\documentclass{IEEEtran}
\usepackage{amsmath}
\usepackage{amssymb}
\usepackage{graphicx}
\usepackage{cite}
\usepackage{fancyhdr}

\graphicspath{{./Figures/}} 
\DeclareGraphicsExtensions{.pdf,.jpg,.png,.jpeg}
   
\newcommand{\overbar}[1]{\mkern 3mu\overline{\mkern-3mu#1\mkern-.0mu}\mkern .0mu}

\begin{document}
	
\title{Models, metrics, and their formulas for typical electric power system resilience events}

\author{ Ian Dobson, Iowa State University
\thanks{Ian Dobson is with the Department of Electrical and Computer Engineering, Iowa State University, Ames Iowa USA; email: dobson@iastate.edu. \newline Support from USA NSF grant 2153163 is gratefully acknowledged.}
}

\fancyhead[c]{\textnormal{ preprint to appear in IEEE Transactions on Power Systems, accepted July 2023. doi: 10.1109/TPWRS.2023.3300125}}
\renewcommand{\headrulewidth}{0.0pt}
\fancyfoot[c]{\textnormal{\small This work is licensed under a Creative Commons Attribution 4.0 License; see http://creativecommons.org/licenses/by/4.0/}}

\maketitle	
\thispagestyle{fancy}

\begin{abstract}
Poisson process models are defined in terms of their rates for outage and restore processes in power system resilience events. These outage and restore processes easily yield the performance curves that track the evolution of resilience events, and the area, nadir, and duration of the performance curves are standard resilience metrics.  This letter analyzes  typical resilience events by 
analyzing the area, nadir, and duration of  mean  performance curves.
Explicit and intuitive formulas for these metrics are derived in terms of the Poisson process model parameters, and these parameters can be estimated from utility data. 
This clarifies the calculation of metrics of typical resilience events, and shows what they depend on. The metric formulas are derived with lognormal, exponential, or constant rates of restoration. 
The method is illustrated with a typical North American transmission event.
Similarly nice formulas are obtained for the area metric for empirical power system data. 
\end{abstract}


\section{\!\!Modeling resilience processes \!with Poisson rates\!}

A resilience event is when many outages bunch up due to stress from extremes such as bad weather.
Performance curves track the progress in time of outages and restores during a resilience event as shown by $\overbar P(t)$ in Fig.~1.
The dimensions of these performance curves are standard metrics of resilience \cite{CarringtonPS21,DobsonPS23,StankovicPS22,PanteliProcIEEE17,NanRESS17}.
Recent research reveals practical stochastic models of the outages and restores and the resulting performance curves in transmission \cite{DobsonPS23} and distribution systems \cite{WeiAM16,CarringtonPS21}.
These models are Poisson processes and their parameters can be estimated from standard data recorded by utilities \cite{CarringtonPS21,DobsonPS23,WeiAM16}.
The mean values of these stochastic models describe the evolution of typical resilience events.
This letter derives formulas for the  area, nadir and duration  metrics describing the dimensions of the mean performance curve. This gives explicit formulas for the metrics of typical resilience events in terms of parameters that can be estimated from observed data. The formulas for area of the performance curve are particularly insightful. 
Of course actual resilience events (realizations of the Poisson processes) show variability about their mean behavior, 
but the mean behavior is useful in describing a typical behavior.
Please refer to \cite{DobsonPS23,StankovicPS22,CarringtonPS21,PanteliProcIEEE17,NanRESS17} for further background and literature review.

\looseness=-1
The outages are modeled as occurring in a Poisson process of rate $\lambda_O(t)$, varying with time $t$. 
 The outages occur in the time interval $[0,o_b]$ and $\lambda_O(t)$ is zero outside the interval $[0,o_b]$.
 The restores are modeled as occurring in a Poisson process of rate $\lambda_R(t)$ in the time interval $[r_a,r_b]$. $\lambda_R(t)$ is zero outside the interval $[r_a,r_b]$. $r_a\ge0$ and $r_b$ can be $\infty$.
It is assumed that there are $n$ outages and $n$ restores in the event. Given the $n$ outages and $n$ restores, the outages and restores are distributed in time according to the probability distributions $\lambda_O(t)/n$ and $\lambda_R(t)/n$ respectively\cite{DobsonPS23}.
One difference with the Poisson models in \cite{DobsonPS23} is that \cite{DobsonPS23} in {\sl extracting} an outage or restore model from the data needs to define the start time of the process with an initial outage or restore, causing an initial delta function in the Poisson rate, whereas here when {\sl applying} a Poisson outage or restore model, one can fix start times for the outage and restore processes ($0$ and $r_a$ respectively) and then assume the outage or restore rate. This simplifies all the formulas.

\begin{figure}[t]
\centering
\includegraphics[width=1.0\columnwidth]{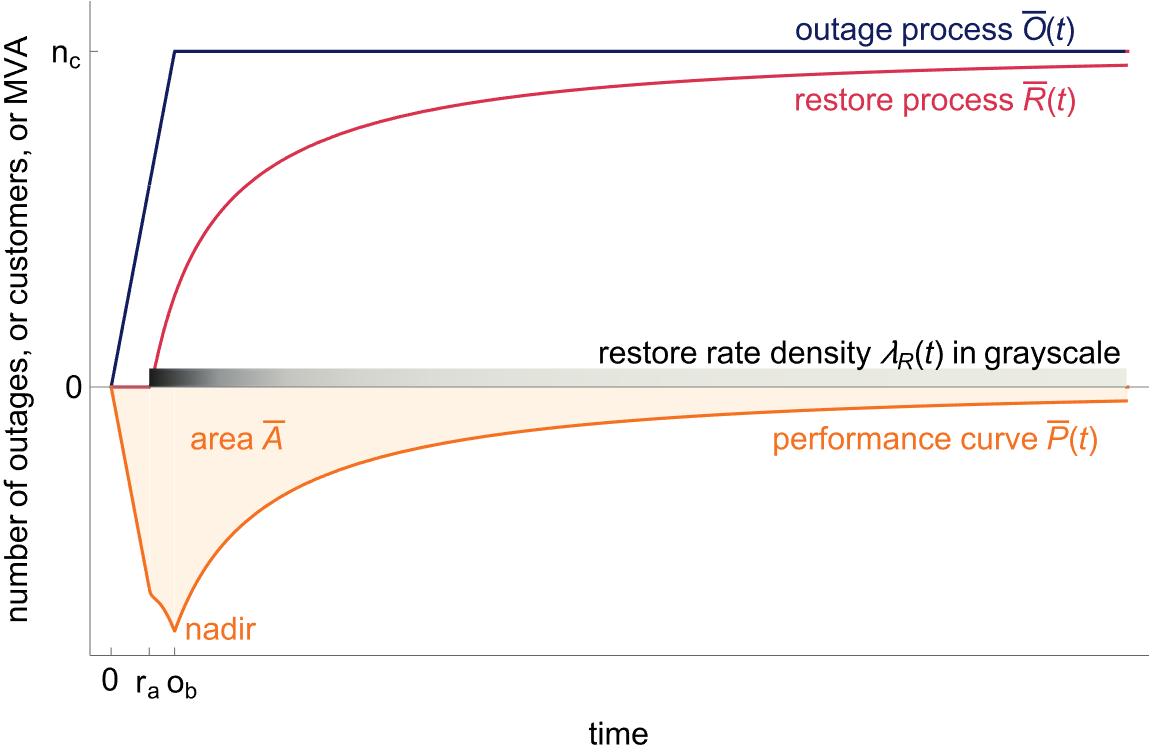}
\caption{Mean processes for a resilience event with constant rate outages in $[0,o_n]$ and restores for $t\ge r_a$ at a slowing, lognormal rate $\lambda_R(t)$. 
The vertical axis lists three different quantities that can be used to track the event processes.
Some of the parameters used are not typical to allow room for labeling; see Fig.~\ref{relatefig2} for a typical example.
}
\label{relatefig}
\end{figure}

The mean cumulative number of outages 
and the  
mean cumulative number of restores at time $t$ are
\begin{align}
\overbar O(t)=\int_{0}^{t}\lambda_O(\tau)d\tau\mbox{ and }\overbar R(t)=\int_{0}^{t}\lambda_R(\tau)d\tau
\label{Ot}
\end{align}

The mean outage and restore rates $\lambda_O(t)$, $\lambda_R(t)$ and mean cumulative outages and restores $\overbar O(t)$, $\overbar R(t)$ easily generalize to track 
outages and restores of other quantities such as customers in a distribution system \cite{CarringtonPS21} or MVA ratings of lines in a transmission system \cite{EkishevaPMAPS22}.
For example, $\lambda_O(t)$ can be the mean rate at which customers outage and  $\overbar O(t)$ can be   the 
mean cumulative customers outaged. These choices change the vertical axis on which 
$\overbar O(t)$, $\overbar R(t)$, $\overbar P(t)$ are plotted.
The mean performance curve $\overbar P(t)$ is the negative of the  mean unrestored amount of the quantity tracked:
\begin{align}
\overbar P(t)=\overbar R(t)-\overbar O(t)
\label{Pt}
\end{align}
At the start of the event, time $t=0$ and 
$\overbar P(0) =0$. $\overbar P(t)$ becomes negative during the event as shown in Fig.~1. Eventually, at time $\infty$, all the outages are restored, $\overbar P(\infty)=0$ and $\overbar O(\infty)=\overbar R(\infty)=n_c$.
Depending on which quantity is tracked, $n_c$ is the total number of outages $n$, the total number of customers outaged, or the total MVA of lines outaged. 
Since $n_c=\overbar O(\infty)=\int_{0}^{\infty}\lambda_O(\tau)d\tau=\overbar R(\infty)=\int_{0}^{\infty}\lambda_R(\tau)d\tau$, $f_o(t)=\lambda_O(t)/n_c$ and $f_r(t)=\lambda_R(t)/n_c$ integrate to one, and are probability distributions of the outage times and restore times respectively. Write $\overline o=\int_0^\infty \tau f_o(\tau)d\tau$ for the mean outage time and $\overline r=\int_0^\infty \tau f_r(\tau)d\tau$ for the mean restore time.

\section{Area and nadir metrics}
\label{general}

This section starts by giving general formulas for the area of the mean performance curve.
This area (regarded as a positive area by including the minus sign)  is
\begin{align}
\overbar A=-\int_{0}^{\infty} \overbar P(t)dt=\int_{0}^{\infty} [\overbar O(t)-\overbar R(t)]dt
\label{A}
\end{align}
Moreover, $\overbar A$ is also  the mean of the area $A$ of the performance curve $P(t)$, since 
\begin{align}\overbar A=-\int_0^\infty {\rm E}P(t)dt=-{\rm E}\int_0^\infty P(t)dt={\rm E}A
\end{align}

Integrating (\ref{A})  by parts, and using  $\overbar P(0)=\overbar P(\infty)=0$ and $\lambda_O(t)=\overbar {O'\!}(t)$, $\lambda_R(t)=\overbar {R'\!}(t)$ gives the very nice formulas
\begin{align}
\overbar A&=\!-\!\int_{0}^{\infty}\hspace{-3mm}t{\overbar{P}'\!}(t)dt=\int_{0}^{\infty} \hspace{-3mm}[t\lambda_R(t)-t\lambda_O(t)]dt={\rm E}[\lambda_R]-{\rm E}[\lambda_O]\notag\\&=n_c\int_{0}^{\infty} \hspace{-3mm}[tf_r(t)-tf_o(t)]dt=n_c(\overline r-\overline o)
\label{generalarea}
\end{align}
According to (\ref{A}), $\overbar A$ is also the area between the mean outage and mean restore processes (see~Fig.~\ref{relatefig2}), so that  (\ref{generalarea}) can be understood as the height $n_c$ of this area  times its average width.

This section now defines the nadir and observes where it occurs.
The nadir $\overbar N$ of $\overbar P(t)$ is the maximum mean number of elements simultaneously outaged during the event, or the negative of the minimum value of $\overbar P(t)$:
\begin{align}
\overbar N=-\min\{\overbar P(t),\ t\ge 0\}
\end{align}

Simulations and frameworks of resilience \cite{StankovicPS22,PanteliProcIEEE17,NanRESS17}
 often make the idealization that outages end before the restores start so that $r_a\ge o_b$.
Then $\overbar P(t)$ is decreasing for  $t<o_b$, constant in $[o_b,r_a]$, and increasing for $t>r_a$.
Therefore the nadir 
occurs at all the times in $[o_b,r_a]$ and is simply $\overbar N=n_c$.
For example, when $\overbar P(t)$ is a trapezoid, the nadir occurs  along the bottom of the trapezoid.
However, in real data \cite{WeiAM16,CarringtonPS21,DobsonPS23,EkishevaPMAPS22}  the restores usually start before the end of the outages. 
Accordingly,  the next paragraph and section~\ref{cases} assume that $r_a< o_b$.

Since  $\overbar R(t)=0$ and $\overbar O(t)$ is increasing for $t<r_a$, and
$\overbar O(t)=n_c$ and $\overbar R(t)$ is increasing for $t>o_b$, the nadir $\overbar N$ of $\overbar P(t)=\overbar R(t)-\overbar O(t)$ must occur inside the time interval $[r_a,o_b]$ or at its endpoints.
$\overbar P(t)$ is a smooth function in $[r_a,o_b]$.

\section{ Metric formulas for different cases}
\label{cases}

\subsection{Constant rate outage and restore processes}

The simplest model has outage  and restore processes 
with constant rates 
$\lambda_O$ and 
$\lambda_R$ respectively.
Then
\begin{align}
\!\!\lambda_O&=\frac{n_c}{o_b} \mbox{ and }
   \overbar O(t)=
\lambda_O t, \ 0\le t< o_b  \label{averageO}\\
\!\!\lambda_R&=\frac{n_c}{r_b-r_a}  \mbox{ and }    \overbar R(t)=
\lambda_R (t-r_a), \ r_a\le t\le r_b
   \label{averageR}
\end{align}
The mean area $\overbar A$ is obtained using (\ref{generalarea}) with the mean restore time $\overline r=\tfrac{1}{2}(r_a+r_b)$ and mean outage time $\overline o=\tfrac{1}{2}o_b$:
\begin{align}
\overbar A&= n_c\, (\tfrac{1}{2}(r_a+r_b)-\tfrac{1}{2}o_b)
\label{areaRconstant}
\end{align}

To compute the nadir $\overbar N$, which always occurs in $[r_a,o_b]$, note that
$\overbar {P'\!}(t)= \lambda_R-\lambda_O$ in $[r_a,o_b]$.
Therefore, if $\lambda_R>\lambda_O$ then $\overbar {P'\!}(t)> 0$  and  $\overbar N=\lambda_O r_a=n_c r_a/o_b$ occurs at time $r_a$. Or if
$\lambda_R< \lambda_O$ then $\overbar {P'\!}(t)< 0$ and $\overbar N=n_c-\lambda_R(o_b-r_a)=n_c(r_b-o_b)/(r_b-r_a)$  
occurs at time $o_b$.

The restore duration can be measured by 
$\overbar D_n=r_b-r_a$.

\subsection{Lognormal rate restore process}

Consider the case of constant rate outage process   (\ref{averageO})
and restore process $\lambda_R(t)$ proportional to lognormal with parameters $\mu$ and $\sigma$ that is used to model typical
North American transmission events in \cite{DobsonPS23}, so that 
$\lambda_R(t)=n_cf_r(t)$ has
\begin{align}
f_r(t)&
=((t-r_a)\sigma \sqrt{2\pi})^{-1}
    \exp[ -(\ln (t-r_a)-\mu)^2/(2\sigma^2)]\notag\\
   \overbar R(t) &=n_c\,\Phi[ (\ln (t-r_a)-\mu)/\sigma]
 ,\qquad\qquad t\ge r_a
   \label{averageRlognormal}
\end{align}
$\Phi$ is the CDF of the standard normal distribution.
The lognormal distribution $f_r(t)$ in (\ref{averageRlognormal}) has mean $\overline r=r_a+\exp[\mu+\sigma^2/2]$
and the area (\ref{generalarea}) becomes 
\begin{align}
\overbar A&= n_c\, (r_a+e^{\mu+\sigma^2/2}-o_b/2)
\label{areaRlognormal}
\end{align}

To compute the nadir, which is  in $[r_a,o_b]$, note that the CDF of the lognormal distribution and hence $\overbar{R}(t)$ and $\overbar{P}(t)$ 
are convex in $[r_a,t_{\rm inflect}]$ and concave for $t\ge t_{\rm inflect}$, where  $t_{\rm inflect}=e^{\mu-\sigma^2}+r_a$ is the mode (maximum) of the lognormal distribution for which $\overbar{R''\!\!\!}\,\,(t_{\rm inflect})=\overbar{P''\!\!\!}\,\,(t_{\rm inflect})=0$. Then
\begin{align}
\overbar {P'\!}(t_{\rm inflect})&=n_c( \sigma\sqrt{ 2\pi})^{-1}
    \exp[-\mu+\sigma^2/2]
-n_c/o_b
\end{align}
Write $t_*$ for a local minimum of $\overbar {P}(t)$. Then  
 $0=\overbar {P'\!}(t_*)= n_cf_r(t_*)-n_c/o_b$, 
and some algebra 
yields a quadratic equation
$[\ln (t_*-r_a)]^2+2(\sigma^2-\mu)[\ln (t_*-r_a)] +\mu^2+2\sigma^2 \ln[\sigma\smash{\sqrt{2\pi}/o_b}]=0$
that can be solved to give  
\begin{align}
\hspace{-2mm} 
t_*&=r_a+\exp\!\big[\mu-\sigma^2-\sigma\sqrt{\sigma^2-2\mu-2\ln[\sigma\sqrt{2\pi}/o_b]}\, \big]\hspace{-1.5mm}
\label{tNlognormal0}\\
&=r_a+\exp\!\big[\mu-\sigma^2-\sigma\sqrt{2\ln[1\!+\!(o_b/n_c)\overbar {P'\!}(t_{\rm inflect})]}\,\big]\hspace{-1.5mm}
\label{tNlognormal}
\end{align}
The negative sign for the square root in (\ref{tNlognormal0}) is chosen to obtain the solution in the convex part of $\overbar {P}(t)$ that 
can be the minimum.
A real solution for $t_*$ exists if $\overbar {P'\!}(t_{\rm inflect})\ge0$.

If $t_{\rm inflect}\le o_b$, then  $\overbar {P'\!}(r_a)=-\lambda_O<0$ and $\overbar{P''\!\!\!}\,\,(t)\ge 0$ in $[r_a,t_{\rm inflect}]$ due to convexity. 
If $\overbar {P'\!}(t_{\rm inflect})\le 0$, then, since $\overbar{P''\!\!\!}\,\,(t)\le  0$ in $[t_{\rm inflect},o_b]$,  $\overbar {P'\!}(t)\le0$ in   
$[r_a,o_b]$ and the nadir occurs at $o_b$.
If $\overbar {P'\!}(t_{\rm inflect})> 0$, the minimum of $\overbar{P}(t)$ in $[r_a,t_{\rm inflect}]$ occurs at $t_*$ and does not occur at 
$t_{\rm inflect}$.
Since $\overbar{P''\!\!\!}\,\,(t)\le 0$ in $[t_{\rm inflect},o_b]$, the minimum of $\overbar{P}(t)$ in $[t_{\rm inflect},o_b]$ could occur at $o_b$. So the nadir occurs at $t_*$ or $o_b$.

\looseness=-1
If $t_{\rm inflect}>o_b$, then reasoning on $[r_a,o_b]$ similar to the reasoning for $t_{\rm inflect}<o_b$ on $[r_a,t_{\rm inflect}]$ shows that 
if $\overbar {P'\!}(o_b)\le 0$, the nadir occurs at $o_b$, and 
if $\overbar {P'\!}(o_b)>0$, the nadir occurs at $t_*$.

One concludes that the nadir occurs at $t_*$ (if it exists in $[r_a,o_b]$) or at $o_b$ so that
$\overbar N=-\min\{P(t_*),P(o_b)\}$.

The duration to reach 95\% restoration can be measured by $\overbar D_{95\%}^{\rm ln}={\rm exp}[\mu+ \sigma\,\Phi^{-1}(0.95)]$. The geometric mean and  median of the positive restore times relative to $r_a$  is $\overbar D_{\rm GM}=e^\mu$  \cite{DobsonPS23}.

\subsection{Exponential rate restore process}
\looseness=-1
Another case  is exponential recovery with time constant and mean restore time $\tau$:
\begin{align}
f_r(t)
    &=\tau^{-1}\, e^{- (t-r_a)/\tau} \notag\\
  \overbar  R(t) &=n_c[1-e^{- (t-r_a)/\tau}], \quad t\ge r_a
  \label{averageRexponential}
\end{align}
The  area (\ref{generalarea}) becomes 
\begin{align}
\overbar A&= n_c\, (r_a+\tau-o_b/2)
\label{areaRexp}
\end{align}

$\overbar R(t)$ is concave, so the nadir occurs at either $r_a$ or $o_b$ and is 
$\overbar N=-\min\{-\lambda_O r_a,n_c[1-e^{- (o_b-r_a)/\tau}]-\lambda_O o_b\}$
$=n_c\max\{r_a/o_b,e^{- (o_b-r_a)/\tau}\}$.

The duration to reach 95\% restoration is $\overbar D_{95\%}^{\rm exp}=\tau\ln 20$.

\section{Area $A$ for empirical data}
\looseness=-1
Similar methods apply to analyzing the data recorded from an actual event, except that $O(t)$, $R(t)$, $P(t)$
are now actual instead of their means. This section analyzes the area $A$ of $P(t)$.
Suppose  the outages occur at times $o_1\le  ... \le o_n$, with quantities $c_1, ..., c_n$. Restores 
 occur at $r_1\le ... \le r_n$, with quantities $c_{\pi(1)},  ..., c_{\pi(n)}$, where $\pi$ is the permutation of $1,2, ...,n$ indicating the order in which the outages restore.
 The  restores in the order of their outage are $r_{\pi^{-1}(1)}, ..., r_{\pi^{-1}(n)}$
 where $\pi^{-1}$ is the inverse permutation\footnote{
 For example, outages $o_1,o_2,o_3$ could restore in the order $o_{\pi(1)}, o_{\pi(2)},o_{\pi(3)}=o_2,o_3,o_1$ where $\pi=\{1\rightarrow2,2\rightarrow3,3\rightarrow1\}$.  $\pi^{-1}=\{2\rightarrow1,3\rightarrow2,1\rightarrow3\}$ and 
 the restores in the order of their outage are $r_{\pi^{-1}(1)}, r_{\pi^{-1}(2)}, r_{\pi^{-1}(3)}=r_3,r_1,r_2$.}.
 
The rates are proportional to delta functions at the outage and restore times, and 
$O(t)$ and $R(t)$ become step functions:
\begin{flalign}
\lambda_O(t)&=\sum_{i=1}^n c_i\delta(t-o_i) \mbox{ and }
O(t)=\sum_{k: o_k\le t} c_k&
\label{cumulativeoutages}\\
\lambda_R(t)&=\sum_{i=1}^n c_{\pi(i)}\delta(t-r_i) \mbox{ and }
R(t)=\sum_{k: r_k\le t} c_{\pi(k)}&\\
\mbox{Then }\ A&={\rm E}[\lambda_R]-{\rm E}[\lambda_O]=\sum_{i=1}^n (c_{\pi(i)} r_i-c_i o_i)&
\label{Adiscrete}
\end{flalign}
Note that permuting the order of the restores in the sum in (\ref{Adiscrete}) makes no difference.
Therefore if one writes $\rho_i=r_{\pi^{-1}(i)}-o_i$ for the time to restore or repair outage $i$, one  obtains
\begin{align}
A&=\sum_{i=1}^n c_i(r_{\pi^{-1}(i)} -o_i)=\sum_{i=1}^n c_i  \rho_i
\label{Adiscreterepair}
\end{align}

A restricted but useful case is now analyzed:
Write $A_x$ for the area when $c_1=c_2=...=c_n=x$ where $x$ is a constant.
For example,  $A_1$ is the area of the performance curve when it tracks the number of outages, and $A_{\overline c}$ is the area of the performance curve when it tracks the number of customers out and one makes the approximation that the average number of customers $\overline c=n_c/n$ is out at each outage so that $c_i=\overline c$.
Equations (\ref{Adiscrete}) and (\ref{Adiscreterepair})  reduce to 
\begin{align}
A_{\overline c}&={\overline c}\sum_{i=1}^n  (r_i-o_i)=n\overline c(\overline r-\overline o)=n_c(\overline r-\overline o)=n_c\, \overbar \rho
\label{A1discrete}
\end{align}
where $\overbar\rho$ is the mean time to repair in the event. Equation (\ref{A1discrete}) recovers the useful (\ref{generalarea}) 
for empirical data in the case that the quantities outaged are interchangeable, including $A_1$.

Another approach \cite{CarringtonPS21} assumes $c_1,c_2,...,c_n$ are sampled from a random variable independent of the outage  and restore times with mean $\overline c$. Then taking the expectation of 
(\ref{Adiscreterepair}) gives  
\begin{align}
\overbar A={\rm E}A=\sum_{i=1}^n {\rm E}c_i{\rm E}[r_{\pi^{-1}(i)} -o_i]=n\overline c(\overline r-\overline o)=n\overline c\,\overline\rho
\end{align}

\section{Typical resilience event example}

\begin{figure}[h]
\centering
\includegraphics[width=1.0\columnwidth]{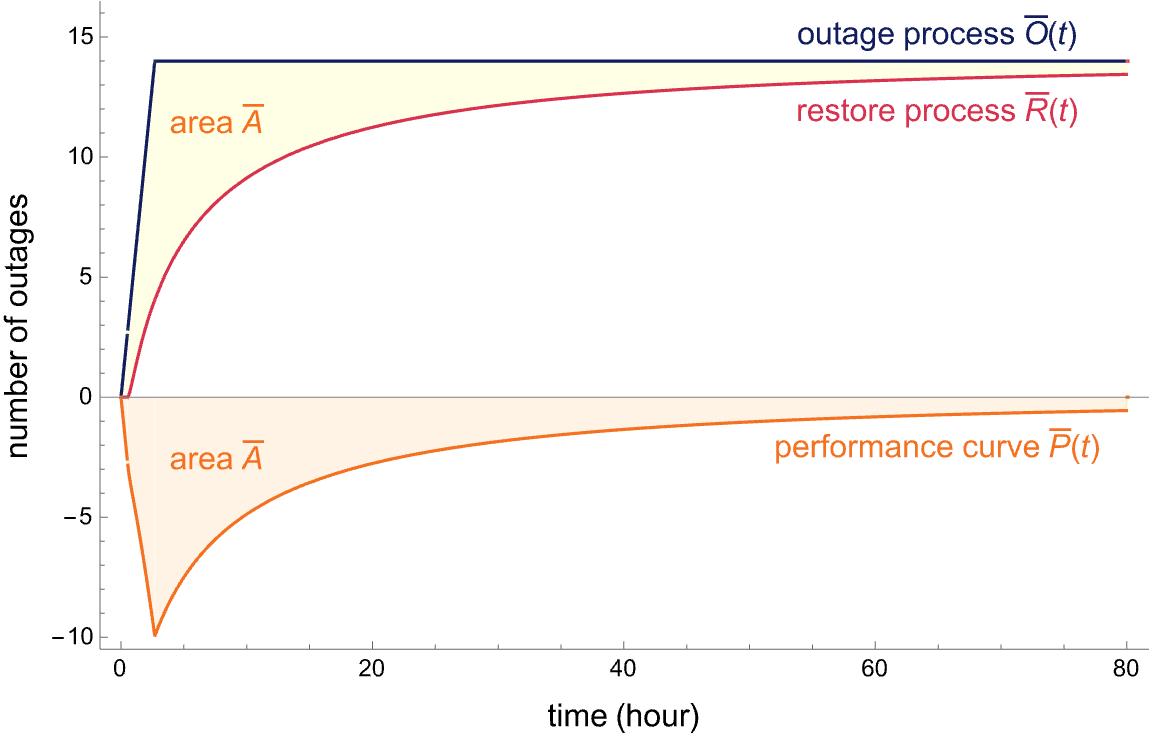}
\caption{ Typical North American transmission system resilience event tracked by number of outages on the vertical scale.
The shaded area $\overbar A$ of $\overbar P(t)$ is the same as the shaded area between 
 $\overbar O(t)$ and $\overbar R(t)$.}
\label{relatefig2}
\end{figure}

This section shows a typical North American transmission resilience event in Fig.~\ref{relatefig2} based on data from \cite{DobsonPS23}.
The event is typical in the class of resilience events with at least 10 outages 
in the bulk  electric power system operated at 100 kV and higher across the continental USA and Canada from 2015 to 2021.
The outage data is collected by the North American Electric Reliability Corporation (NERC) in their Transmission Availability Data System (TADS).
Some applications would usefully consider the events typical for subsets of data such as events with
specific causes such as hurricanes or winter storms, or events in particular seasons or regions, but this example 
considers all the data.
Fig.~\ref{relatefig2} uses the median parameters for these event data from \cite{DobsonPS23}:
number of outages $n = 13.5$, which is rounded up to $n=14$,
outage duration $o_b = 2.69$ h,
time to first restore $r_a = 0.52$ h, and
lognormal distribution restoration parameters $\mu =1.64$ and
$\sigma = 1.56$ 
($\mu =\ln{D_{\rm GM}}$ is calculated with $D_{\rm GM}=5.15$ h).

\section{Discussion and conclusions}

\looseness=-1
This letter introduces Poisson process models in terms of their rates for outage  and restore  resilience processes (\ref{averageO},\ref{averageR},\ref{averageRlognormal},\ref{averageRexponential}) and performance curves (\ref{Pt}). The model parameters can be 
calculated from standard utility outage data \cite{DobsonPS23}.
One of the parameters $n_c$ is the event total of the quantity tracked by the processes, such as  number of outages, or number of customers, or MVA rating of the lines.
Section~\ref{cases} gives explicit formulas in terms of the parameters for area, nadir, and duration metrics describing the mean performance curve $\overbar P(t)$ for a constant rate outage process and constant, lognormal, or exponential rate restore processes.
In each case, the restore duration has a different definition and formula, and
the event duration is always the restore duration plus the time to the first restore $r_a$.
In all these cases, the nadir of $\overbar P(t)$ is proportional to $n_c$ and occurs only at the end of the outages  $o_b$ or at another time that can be explicitly calculated. 
Then the nadir is calculated as 
the minimum of $\overbar P(t)$  at the two times. Moreover, for the  North American transmission events  in  \cite{DobsonPS23}, 98\% of the nadirs  occur at the end of the outages so that $\overbar N=-\overbar P(o_b)$.

\looseness=-1
$\overbar A$ is area of the mean performance curve, area between the mean outage and restore processes, and expected area of the performance curve.
$\overbar A$ is given by
 (\ref{generalarea},\ref{areaRconstant},\ref{areaRlognormal},\ref{areaRexp}) as the difference between the average restore and outage rates, or  $n_c$ 
times the difference between the average restore and outage times. 
Thus  10\% reduction in $n_c$ causes the same 10\% reduction of $\overbar A$ as 10\% faster restoration.

The area $A$ of the performance curve for empirical data is also the difference between average restore and outage rates (\ref{Adiscrete}), 
and is expressed in terms of component repair times in (\ref{Adiscreterepair}), which, in the case of interchangeable outages such as when 
counting the number of outages, simplifies to   $n_c$ times the average repair time in (\ref{A1discrete}). This links the outage and restore process systems view with the individual component reliability view.

\looseness=0
Areas of performance curves or mean performance curves represent component hours, customer hours, or MVA hours  in a resilience event, and the usefulness of these 
metrics underlines the importance of giving new derivations for intuitive formulas showing how the areas depend on $n_c$ and average outage, restore, and repair times.

\bibliographystyle{IEEEtran}

\end{document}